\documentclass[letterpaper, 10 pt, conference]{ieeeconf}  

\IEEEoverridecommandlockouts                              
\title{\LARGE \bf
RTOS Architectures that Solve the Diminishing Bandwidth Problem (with Multi-core Support)
}

\author{Mazen Arakji}

\usepackage{csquotes}
\usepackage{xcolor}
\usepackage{booktabs}
\usepackage{adjustbox}
\usepackage{url}
\usepackage{etoolbox}
\let\labelindent\relax
\usepackage{enumitem}
\apptocmd{\sloppy}{\hbadness 10000\relax}{}{}
\begin{document}

\setlength\parindent{0pt}

\maketitle
\thispagestyle{empty}
\pagestyle{empty}

\begin{abstract}

The \textit{Diminishing Bandwidth Problem} is a long standing, previously unidentified, extensibility problem of current real-time operating systems characterized by a superficial dependency between the number of tasks in a system and the maximum bandwidth associated with an unrelated peripheral device.  In the worst case, this diabolical deficiency will continue to decrease the maximum bandwidth of a peripheral device as more tasks are added to the application.  If this is not taken into account, a previously functional application may experience data loss if more tasks are added to it in order to, for example, implement new features.  Three novel RTOS architectures that solve the \textit{Diminishing Bandwidth Problem} are specified and discussed: the \textit{Defer Structure RTOS Architecture}, the \textit{Barriers and Requests RTOS Architecture}, and the \textit{Strictly Atomic RTOS Architecture}.  Finally, two hardware solutions to the \textit{Diminishing Bandwidth Problem} are also presented\footnote{All concepts described here are captured in U.S Patent 11,507,524 and U.S Patent 11,734,051 and other pending claims.  Additional implementation details for those with less than ordinary skill in the development of real-time operating systems are shown.}.
\end{abstract}
\hfill \break
\hfill \break
\hfill \break
\section{Introduction}
Many embedded systems have real-time constraints.  Concretely, upon receiving its input, a system may have to produce its output within a certain deadline.  Current real-time operating systems solve this problem and are therefore a common component of such embedded systems.  Unfortunately, however, they also introduce the possibility of data loss.  If such data loss is observed, either the bandwidth of the incoming data must be decreased, or if this is not under the control of the system developer, the other solution is to utilize a more powerful system with a higher operating frequency.  This is unnecessary.  An even worse scenario is when data is lost unbeknownst to the user - in a scientific application, this is a loss of knowledge. 
\hfill \break
\hfill \break
The paper is organized as follows: Section II presents a background on embedded systems and real-time operating systems.  A basic understanding of developing applications for real-time operating systems is expected.  Section III introduces the \textit{Defer Structure RTOS Architecture} and compares and contrasts four defer structures.  Section IV presents the \textit{Barriers and Requests RTOS Architecture} and describes two different embodiments.  Section V introduces the \textit{Strictly Atomic RTOS Architecture}.  Two hardware solutions to the \textit{Diminishing Bandwidth Problem} are described in Section VI; the first is for systems where a DMA controller is unavailable or unneeded, while the second introduces a new DMA controller design.  Section VII considers multi-core systems and associated implications.  Section VIII contains a discussion with concluding remarks.
\hfill \break
\hfill \break
\hfill \break
\section{Background}
The ability of a computer to receive data from external hardware sources (peripherals) is a fundamental functionality.  In the beginning, computers would simply only ever execute the very next instruction.  Within a loop, the computer would check for new data from all peripherals.  When a particular peripheral had new data available, the computer would process it.  However, it was discovered that such a system could lead to loss of data.  This is true especially as the number of peripherals being monitored increased.  The reason for the data loss is straightforward.  While the computer is busy processing the data from one peripheral, another peripheral might receive multiple data, effectively overwriting the old value before it was processed.
\hfill \break
\hfill \break
The solution to this problem was interrupts [1,2].  This is an actual hardware change to the processor itself.  Concretely, new pins were added.  These interrupt pins, were essentially new input to the processor.  They were each connected to a peripheral.  Upon new data becoming available at a particular peripheral, the interrupt pin connected to that peripheral would be asserted.  This would cause the computer to, instead of executing the very next instruction as usual, begin executing instructions from a predetermined location in memory.  And so, clearly, these instructions (called an interrupt handler) would retrieve the data from the peripheral, store it in a particular buffer in memory set aside for that peripheral for future processing, and then return to the prior, regular, instruction execution.  In this paper, the term \enquote{interrupt} refers exclusively to hardware interrupts of peripherals associated with I/O, and is to be distinguished from software interrupts or timer interrupts (e.g. \enquote*{Sys Tick} timer interrupt or similar).  With this change to the processor, namely, adding interrupts, the problem of data loss was solved.  Thus, in a computer system with interrupts, instead of having the computer check for new data from all peripherals as before, the computer could check its own memory, and specifically the memory buffer associated with each peripheral.  When a memory buffer of a particular peripheral had new data or even multiple data, the computer could process it.
\hfill \break
\hfill \break
In addition to being able to retrieve data from peripherals without loss, many computer systems have requirements on how long it can be before the data that has been received and stored by the interrupt handler finally begins to be processed.  This is the response time.  In the previously described computer system with interrupts, assume that a particular peripheral, peripheral A, is considered to be of the highest priority.  That is, when new data is received and stored by the interrupt handler associated with peripheral A, it must be processed within a specific deadline.  A terrible response time occurs in this scenario:  The computer checks the memory buffer associated with peripheral A and finds that no new data is available.  Then it checks the memory buffer associated with some other, lower priority peripheral, peripheral B, and discovers that it contains new data.  In the same instant that the computer then begins to process this data (task B), new data from peripheral A arrives.  The interrupt pin connected to peripheral A is asserted and the interrupt handler for peripheral A executes, retrieving the data from peripheral A and storing it in the memory buffer associated with peripheral A.  The interrupt handler then returns to the prior, regular, instruction execution, namely, task B.  Unfortunately, of all the peripherals, the time it takes to process the data from peripheral B is the longest.  Finally, upon completing task B, the system can check the memory buffer associated with the high priority peripheral A and process the received data (task A).  This system was forced to wait for the longest task to execute from beginning to end before being able to respond to the new data from the high priority peripheral.
\hfill \break
\hfill \break
The solution to this problem was real-time operating systems, and concretely, the context switch.  The context switch is a concept from the earliest time sharing operating systems [3] and is in fact the defining characteristic of an RTOS.  It was repurposed to solve the response time problem.  A context switch is purely a software solution, and in its simplest form would be most easily described by reexamining the previous example:  As before, in the same instant that the computer begins the processing of peripheral B's data (task B), new data from peripheral A arrives.  And as before, the interrupt handler associated with peripheral A executes.  However, after retrieving the data from peripheral A and storing it in the memory buffer associated with peripheral A, a context switch is executed.  Instead of returning to the prior, regular processing, namely, task B, the context switch saves this for later, and first executes the processing of peripheral A's data (task A).  Thus, the response time for processing peripheral A's data is now constant and essentially equal to the context switch time.  This solves the unbounded worst case response time for processing peripheral A's data.
\hfill \break
\hfill \break
The problem with this solution, however, is that during the context switch, interrupts are disabled to update the internal data structures of the RTOS.  These data structures are typically linked lists used to keep track of tasks that are blocked or ready to run.  There is one general Ready List and a separate Blocked List for each semaphore\footnote{There is also one Delayed List that includes tasks that are waiting on a timeout.  This list sorted by timeout time and when a timeout occurs, the \enquote*{Sys Tick} interrupt handler executes to remove the task from the Delayed List and insert it into the Ready List.  Tasks that are blocked that want to block only for some maximum amount of time would be in both a Blocked List and the Delayed List simultaneously.}$^{,}$\footnote{Semaphores are locks that multiple tasks can block on until the lock is released by another task or an interrupt handler (the lock release operation essentially involves extracting the highest priority task from the semaphore's Blocked List and inserting it into the Ready List).  Counting semaphores have multiple locks and tasks will block on them only if all locks are locked (unreleased).}.  These lists are sorted by task priority and can be accessed concurrently by multiple interrupt handlers, the software interrupt handler, the \enquote*{Sys Tick} interrupt handler, or the currently executing task itself.  If interrupts are not disabled while a task is being inserted into or extracted from a Blocked List or the Ready List, then the list may be rendered invalid.  For example, a lower priority interrupt handler my be preempted by a higher priority interrupt handler before the lower priority interrupt handler completes its insertion into the Ready List, leaving the Ready List in an invalid state and preventing the high priority interrupt handler from properly executing its insertion (and thus possibly further corrupting the Ready List).
\hfill \break  
\hfill \break
Disabling interrupts in this way is the main cause of the \textit{Diminishing Bandwidth Problem}.  The reason it is an extensibility problem, is that these lists have a key property: their size is bounded by the number of tasks in the system.  Therefore, the amount of time that interrupts may be disabled increases as more tasks are added to the system.  Consider, for example, that at the time data arrives at a peripheral associated with a (currently blocked) high priority task, the currently executing task is either a low priority task that had just disabled interrupts in order to block, or any other task that had just disabled interrupts in order to delay for a long time.  In the former case, the executing low priority task must be inserted into the priority-sorted linked list of tasks blocked on the same semaphore, and in the latter case, the task must be inserted into the time-sorted linked list of tasks that are delayed.  All such linked list data structures have a sorted insertion running time of O(n).  Therefore, in the worst case, the maximum time interval in which interrupts may be disabled increases linearly with the number of tasks (assuming 8 cycles/iteration of the loop used to insert into the sorted linked list, this comes to 8 cycles/task which is considerable since the length of the entire interrupt handler is only 100-200 cycles).  A linear dependency between the number of tasks in the system and the processor operating frequency required to maintain the maximal bandwidth that can be accepted at some peripheral is thus created.  Therefore, an application designed to process input at some bandwidth, may fail after increasing the number of tasks in the system in order to perhaps enhance the functionality of the application.  Although modern SoCs provide hardware buffering at the peripheral itself (in order to address interrupt latency\footnote{Interrupt latency is a specification of the processor hardware (the value is immutable and can be looked up in the processor data sheet).  It is defined as the number of cycles between when an interrupt is asserted and when the first instruction at the associated interrupt entry point executes.}), this is not sufficient to also solve the \textit{Diminishing Bandwidth Problem}, and as more tasks are added to the system, the maximum bandwidth that can be accepted at some peripheral will continue to decrease.
\hfill \break
\hfill \break
Ultimately, and unequivocally, an RTOS should not internally disable an interrupt that it did not enable and that was enabled by the application developer.
\hfill \break
\hfill \break
\hfill \break
\section{THE DEFERRED STRUCTURE RTOS ARCHITECTURE}
In an RTOS, after data is received at a peripheral, an interrupt handler will retrieve and store this data then release a semaphore (e.g. smphrGvIsr() RTOS function or similar), which will disable interrupts, increment the semaphore, and unblock the task that will process this data (remove it from the semaphore's Blocked List and insert it into the Ready List) and if this task is a higher priority than the currently executing task, a software interrupt (set at the lowest priority along with the \enquote*{Sys Tick} interrupt) is asserted whose handler will switch the lower priority task for the higher priority one.  See Fig. 1.  It is also very important to note that the length of the interrupt handler is significantly increased by the RTOS and this is especially true if the unblocked task must also be removed from the Delay List (not shown in Fig. 1 but possible if the unblocked task was blocked with a timeout), because longer interrupt handlers also diminish bandwidth.
\hfill \break
\hfill \break
The \textit{Deferred Structure RTOS  Architecture} is fundamentally different.  The interrupt handler, after incrementing the semaphore, places a reference to it in a defer structure (assume a FIFO).  Then the lowest priority software interrupt is asserted.  The software interrupt handler will remove the semaphore reference from the defer structure and extract the highest priority task in its Blocked List and insert it into the Ready List (and switch tasks if needed).
\hfill \break
\begin{figure}[h!]
  \centering
    \includegraphics [scale=0.27] {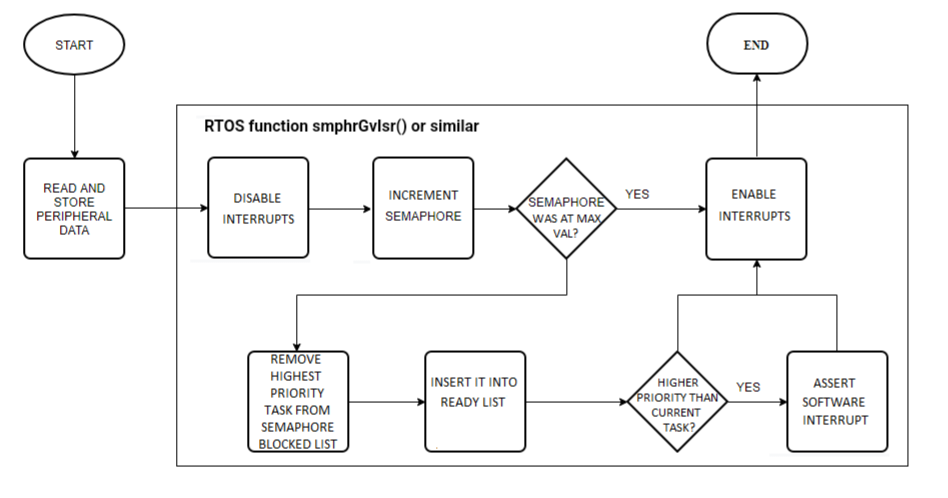}
  \caption{Current RTOS Interrupt Handler}
  \end{figure}
\hfill \break
Since only the software interrupt handler accesses Blocked Lists and the Ready List, it is not possible for interrupt handlers to interfere with one another and  corrupt these lists by accessing them concurrently.  Furthermore, the software interrupt handler can be preempted at any time (whether it is accessing a Blocked List, the Ready List, or reading the defer structure) by any interrupt handler which is free to place a new reference to a semaphore in the defer structure and reassert the software interrupt.  The software interrupt handler, after resuming and running to completion, because its interrupt was reasserted, will run again and then discover the new semaphore reference in the defer structure.  Tasks wishing to access Blocked Lists or the Ready List need only disable the software interrupt to ensure exclusive access to these lists\footnote{Implied here is also disabling the \enquote*{Sys Tick} interrupt since its handler will access the Ready List and possibly a Blocked List.}.
\hfill \break
\hfill \break
Because interrupt handlers no longer access Blocked Lists, the Ready List, or the Delay List, but rather only add a semaphore reference to the defer structure, the length of the interrupt handler remains relatively short.  Nonetheless, this defer structure is being concurrently accessed by interrupt handlers.  Interrupt handlers, however, should not disable interrupts in order to do so.  There is a better solution: The defer structure can be accessed using atomic instructions.  For example ARM architectures contain the load exclusive and store exclusive instructions [4] and x86 architectures include the cmpxchg instruction [5].  Atomic instructions are under-utilized by system developers and application developers, since their use is typically confined to implementing spin locks for multi-core platforms.  A proper RTOS should provide portable functions for atomic operations.  See Fig 2.
\hfill \break
\begin{figure}[h!]
  \centering
    \includegraphics [scale=0.12] {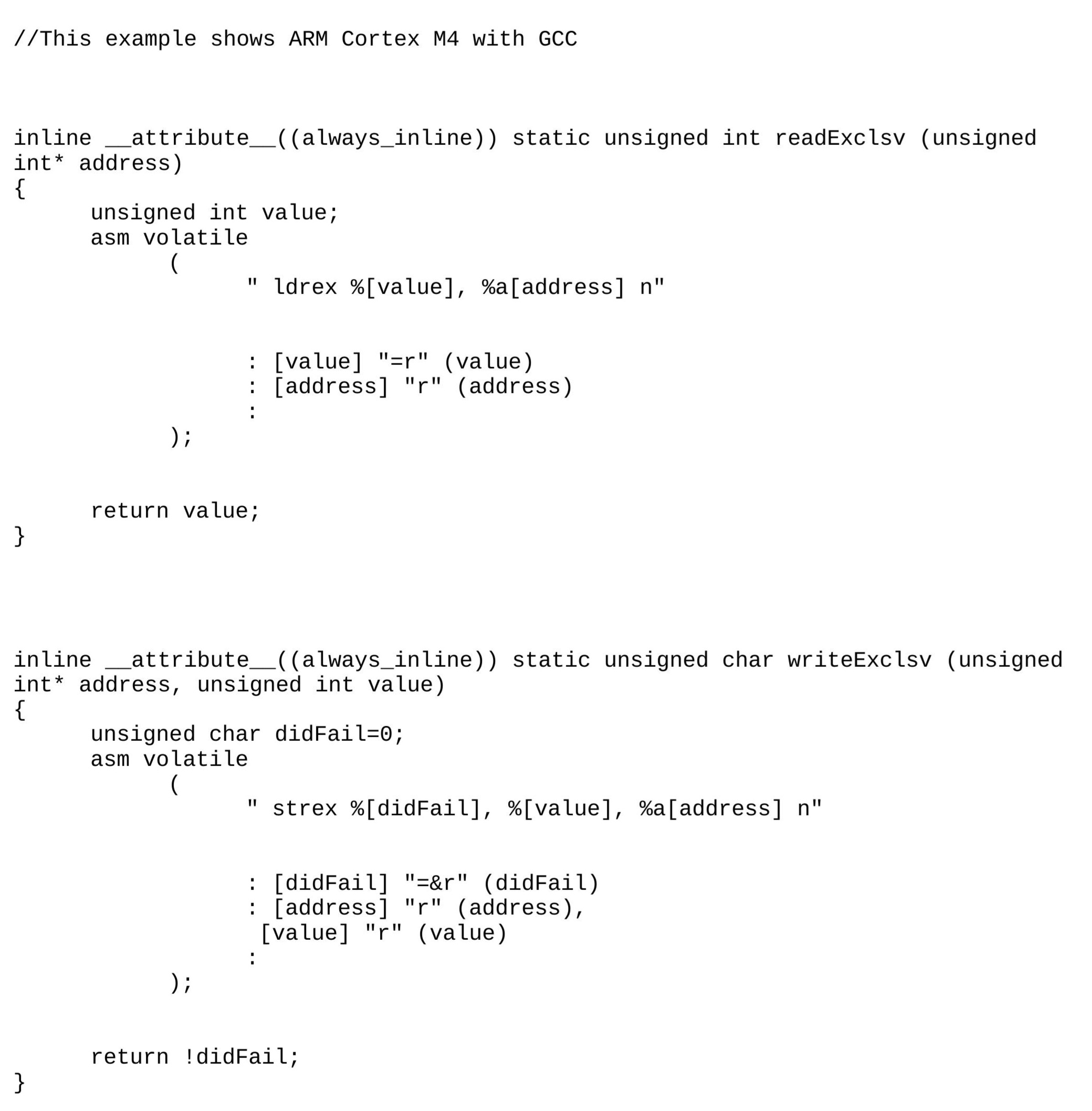}
  \caption{LOAD/STORE Exclusive Functions}
  \end{figure}
\hfill \break
\hfill \break
The exact method by which atomic instructions can be used to access the defer structure requires closer inspection of the defer structure itself.  There are four unique defer structures that will be discussed, where the defer structure is characterized by both its type and maximum size.  In all cases, however, it is clear that the size of the defer structure must be guaranteed to not fail the application (e.g. buffer overflow cannot be possible).  The real-time application developer assumes that the RTOS will be functional and has no idea if or how defer structures are being used.  If buffer overflow is possible, for example when all interrupts occur simultaneously (or when enough interrupts occur consecutively), or when after adding more interrupts to an application (to implement new features) all interrupts occur simultaneously (or enough interrupts occur consecutively), then this effectively introduces a new extensibility problem.  The RTOS will seem to be functional but may fail in the field.  For real-time applications, where human life or significant assets may be at stake, this unacceptable.
\hfill \break
\hfill \break
\subsection{FIFO Semaphore Counts Buffer}
The first type of defer structure is a FIFO implemented with a circular buffer.  Interrupt handlers (in the smphrGvIsr() RTOS function) write semaphore references at the 'head' and the software interrupt handler reads them from the \enquote*{tail}.  If interrupts are neither disabled nor atomic instructions used, corruption of the circular buffer is possible.  For example, consider the case where two interrupt handlers A (high priority) and B (low priority) concurrently attempt to access the FIFO Buffer. Specifically, interrupt handler B executes first, and inserts its semaphore reference as follows:
\hfill \break
\hfill \break
\begin{enumerate}[label={\arabic*.}]
  \item The \enquote*{head} of the FIFO Buffer is read from memory.
  \item The semaphore reference is stored at the location of the \enquote*{head}.
  \item The \enquote*{head} of the FIFO Buffer is incremented.
\end{enumerate}
\hfill \break
If after step 2 executes, interrupt handler B is interrupted by interrupt handler A, then interrupt handler A will copy its semaphore reference to the FIFO Buffer by first reading the \enquote*{head} variable, then storing its semaphore reference at the 'head' location. The \enquote*{head} variable read by interrupt handler A is the same one that was used by interrupt handler B, since interrupt handler B did not yet increment it. Thus, when interrupt handler A stores its semaphore reference at the 'head' location, it will effectively overwrite interrupt handler B's semaphore reference.  This problem is averted by using atomic instructions.  See Fig. 3.
\begin{figure}[h!]
  \centering
    \includegraphics [scale=0.27] {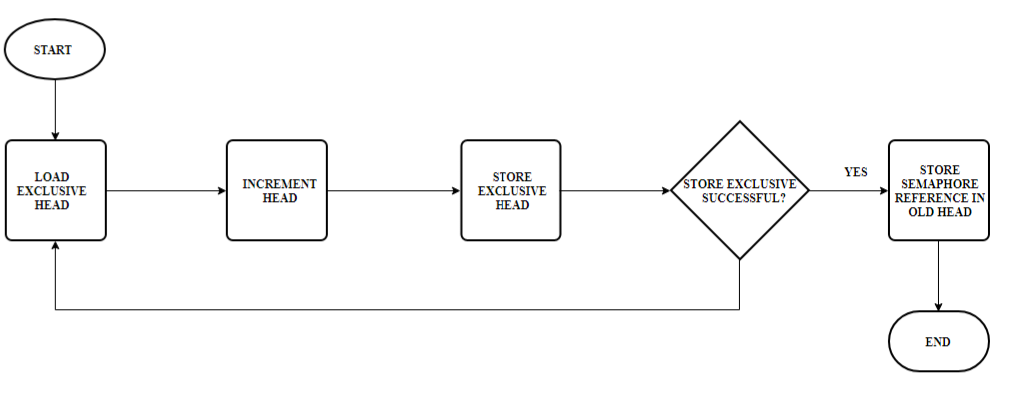}
  \caption{Atomic Instructions to Insert into FIFO Buffer}
  \end{figure}
\hfill \break
\hfill \break
After placing a reference to the semaphore in the FIFO buffer, the first task blocked on the semaphore is checked to see if it is a higher priority than the currently executing task.  If it is, the software interrupt is asserted.  If the currently executing task is a higher priority, then the semaphore reference is left in the FIFO buffer.  See Fig. 4.  When the currently executing task unblocks a task or blocks itself, it will disable the software interrupt (and the \enquote*{Sys Tick} interrupt), process all semaphores in the FIFO buffer, and then if it has unblocked a higher priority task or blocked itself, it will assert the software interrupt before enabling it.  See Fig. 5.  The currently executing task must first process the FIFO buffer to ensure that the order of unblocked tasks is not corrupted.  For example, if an interrupt handler places a reference to a semaphore in the FIFO buffer, but the highest priority task blocked on the semaphore is a lower priority than the currently executing task (software interrupt is not asserted), and at some time much later, the currently executing task unblocks a task of equal priority (to the one blocked on the semaphore in the FIFO buffer), then this task should not be placed into the Ready List first\footnote{The \enquote*{Sys Tick} interrupt handler must also process the FIFO Buffer before placing a task in the Ready List, based on the same line of reasoning.}.
\hfill \break
\begin{figure}[h!]
  \centering
    \includegraphics [scale=0.27] {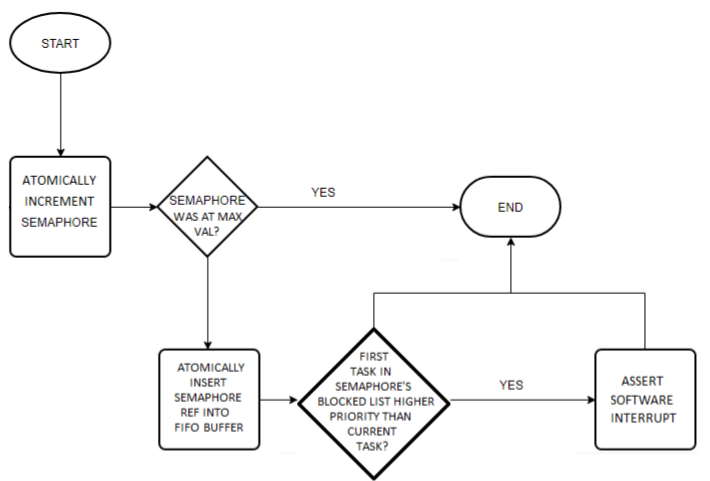}
  \caption{FIFO Semaphore Counts Buffer Interrupt Handler - smphrGvIsr() only}
  \end{figure}
\begin{figure}[h!]
  \centering
    \includegraphics[scale=0.27]{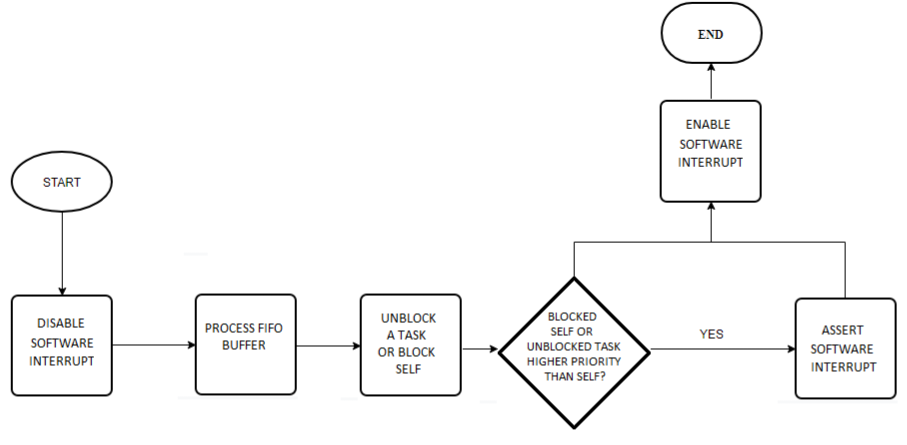}
  \caption{FIFO Semaphore Counts Buffer - Task Blocking Self or Unblocking other Task}
\end{figure}
 \hfill \break
The size of this defer structure is bounded by the number of semaphore counts of semaphores released from interrupt handlers (counting semaphores are a fundamental synchronization primitive of any proper RTOS).  The reason is, anytime a semaphore is incremented, a new reference to it is placed in the FIFO buffer.  In order to ensure that real-time applications do not fail, this number of semaphore counts must be statically specified in a configuration file (e.g. \#define NUM\_ISR\_SEMPHR\_COUNTS).  For a 32 bit system, the size of the FIFO buffer would be NUM\_ISR\_SEMPHR\_COUNTS\textless\textless2.
\hfill \break
\hfill \break
\subsection{FIFO Semaphore Buffer}
The second type of defer structure is also a FIFO implemented with a circular buffer.  However, in this case the size of the circular buffer is bounded only by the number of semaphores released from interrupt handlers (this must also be statically specified in a configuration file).  For the previous defer structure, an application with even only a few semaphores released from interrupt handlers, but that has a high number of semaphore counts for each semaphore, will cause the size of the buffer to become too big for embedded systems.
\hfill \break
\hfill \break
For this defer structure, when a semaphore is incremented (by an interrupt handler), if the buffer already contains a reference to this semaphore, no new reference is added to the buffer.  Rather, an \enquote*{Unblock Count} variable of the semaphore is incremented (only semaphores released from interrupt handlers require the \enquote*{Unblock Count} variable so when a semaphore is created, an additional parameter is required to identify it as such).  See Fig. 6.  The software interrupt handler will unblock \enquote*{Unblock Count} tasks from the semaphore's Blocked List.  Since this \enquote*{Unblock Count} variable may be accessed concurrently by many interrupt handlers and the software interrupt handler, it must be accessed atomically.  An informal (uncompiled) simple example of how the software interrupt handler processes the FIFO buffer is provided in Fig. 7.  Note that when the currently executing task unblocks a task or blocks itself, there is no need to process all semaphores in the FIFO buffer as before (since semaphore references are not left in the FIFO buffer).
\hfill \break
\begin{figure}[h!]
  \centering
    \includegraphics [scale=0.27] {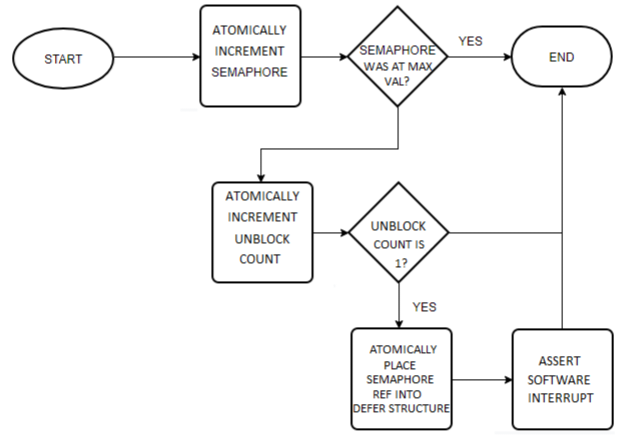}
  \caption{FIFO Semaphore Buffer Interrupt Handler - smphrGvIsr() only (also all Defer Structures that have 'Unblock Count' Variable)}
  \hfill \break
  \end{figure}
\begin{figure}[h!]
  \centering
    \includegraphics[scale=0.12]{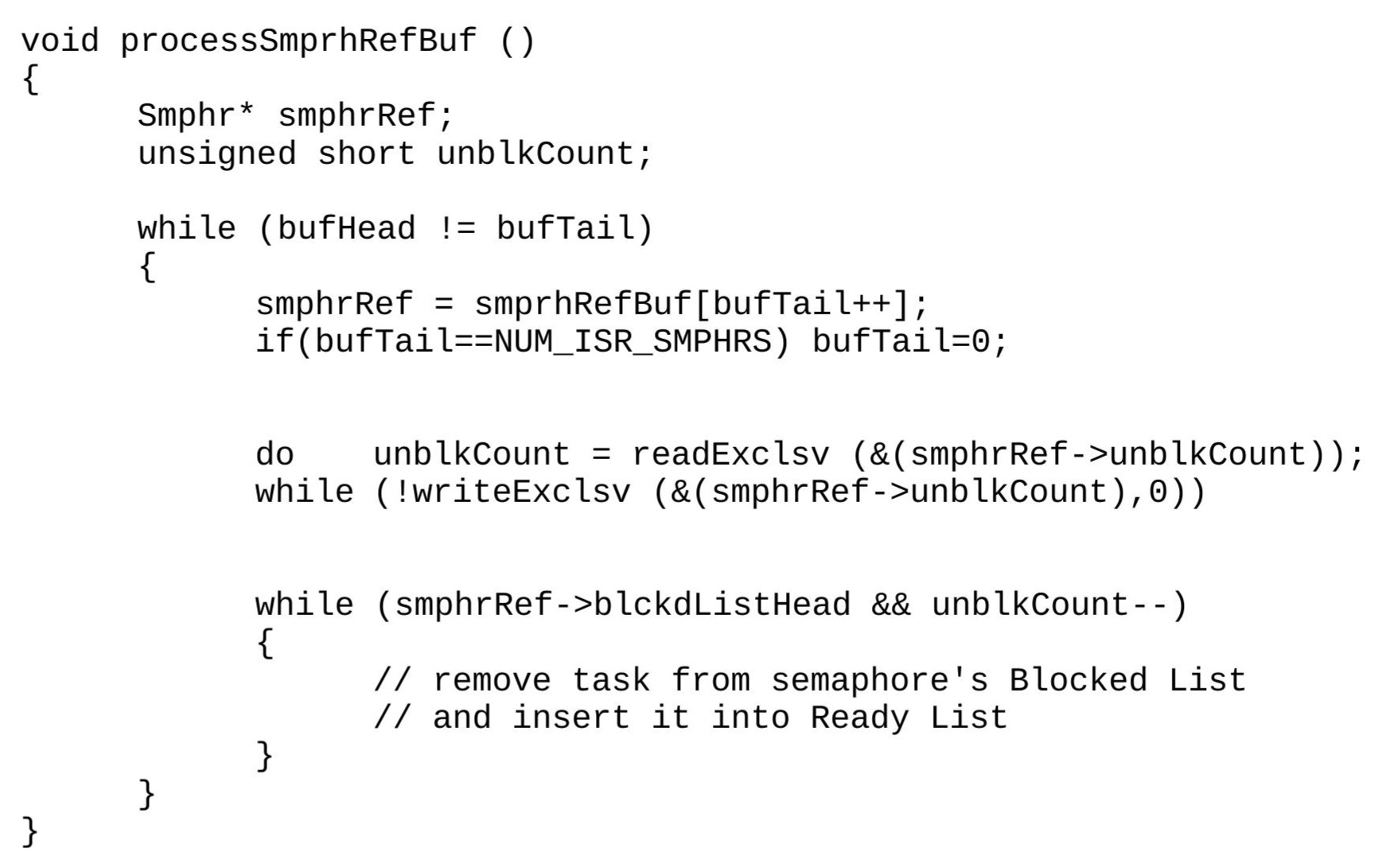}
  \caption{SW Interrupt Handler - Process FIFO Semaphore Buffer}
\end{figure}
\hfill \break
 Note that when an interrupt handler does place a semaphore reference in the FIFO buffer, it always asserts the software interrupt (regardless of whether or not the highest priority task blocked on the semaphore has a priority higher than that of the currently executing task).  The reason for this is to ensure that the order of unblocked tasks is not corrupted.  For example, if a currently executing high priority task executes for a long time, it is possible that when it first begins executing, two interrupts occur and two semaphore references are placed on the FIFO buffer (where the highest priority task blocked on each semaphore is of equal medium priority).  At some time much later, while the high priority task is still executing, the first interrupt occurs again and the 'Unblock Count' variable of the first semaphore is incremented (the second task in the first semaphore's Blocked List is also of equal medium priority).  Now when the highest priority task blocks and the FIFO buffer is processed, the second task in the first semaphore's Blocked List will be incorrectly placed into the Ready List before the first task in the second semaphore's Blocked List.  Always forcing a software interrupt after placing a semaphore reference in the FIFO buffer ensures such corruption is not possible.
\hfill \break
\hfill \break
This defer structure, for some applications, can be much smaller than the previous one (FIFO Semaphore Counts Buffer).  However, every semaphore released from an interrupt handler, must contain an \enquote*{Unblock Count} variable (assume 2 bytes is sufficient). Furthermore, an RTOS based on this defer structure can add significant latency to the execution time of the highest priority task.  The reason is that, every interrupt that occurs, that causes a semaphore reference to be added to the FIFO buffer, also causes the software interrupt to be asserted.  Effectively, the interrupt latency is doubled (note that this latency, is in addition to the latency incurred by buffering and unbuffering, characteristic to both defer structures).  Finally, note that an RTOS based on this defer structure requires both static configuration (the number of semaphores released from interrupt handlers specified in a configuration file) and dynamic configuration (identification of semaphores released from interrupt handlers via an additional parameter during creation in order to create a semaphore that includes the \enquote*{Unblock Count} variable).
\hfill \break
\hfill \break
\subsection{FIFO Linked List}
This defer structure is another FIFO but implemented as a linked list. The only difference is that static configuration is unneeded (the number of semaphores released from interrupt handlers need not be statically specified in a configuration file).  Rather, it is sufficient to, when a semaphore is created at runtime, include an additional parameter to specify that this semaphore will be released from an interrupt handler; thus the semaphore will include both the old \enquote*{Unblock Count} variable and a new \enquote*{next} pointer (4 bytes per semaphore).  It is important to keep the insertion into the FIFO Linked List by interrupt handlers concise.  In order to do so, the FIFO Linked List should be initialized with an unused semaphore (\enquote*{Unblock Count} variable = 0).  This averts the need for interrupt handlers to take initial conditions into account (concretely the \enquote*{head} of the list need never be considered).  See Fig. 8.  An informal (uncompiled) simple example of how the software interrupt handler processes the FIFO Linked List is shown in Fig. 9.
\hfill \break
\begin{figure}[h!]
  \centering
    \includegraphics [scale=0.27] {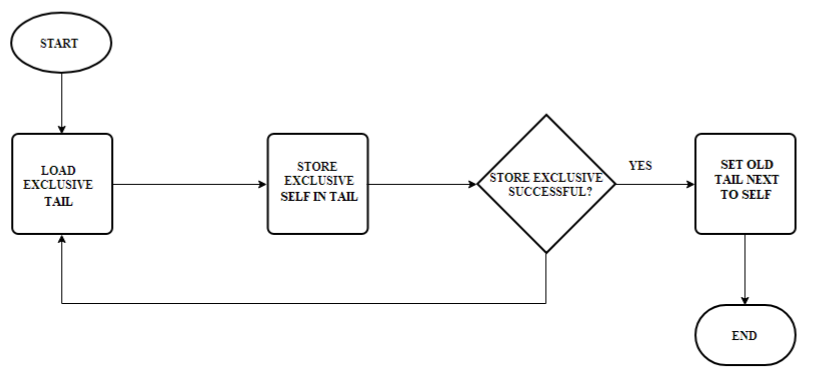}
  \caption{Atomic Instructions to Insert into FIFO Linked List}
  \end{figure}
\begin{figure}[h!]
  \centering
    \includegraphics[scale=0.12]{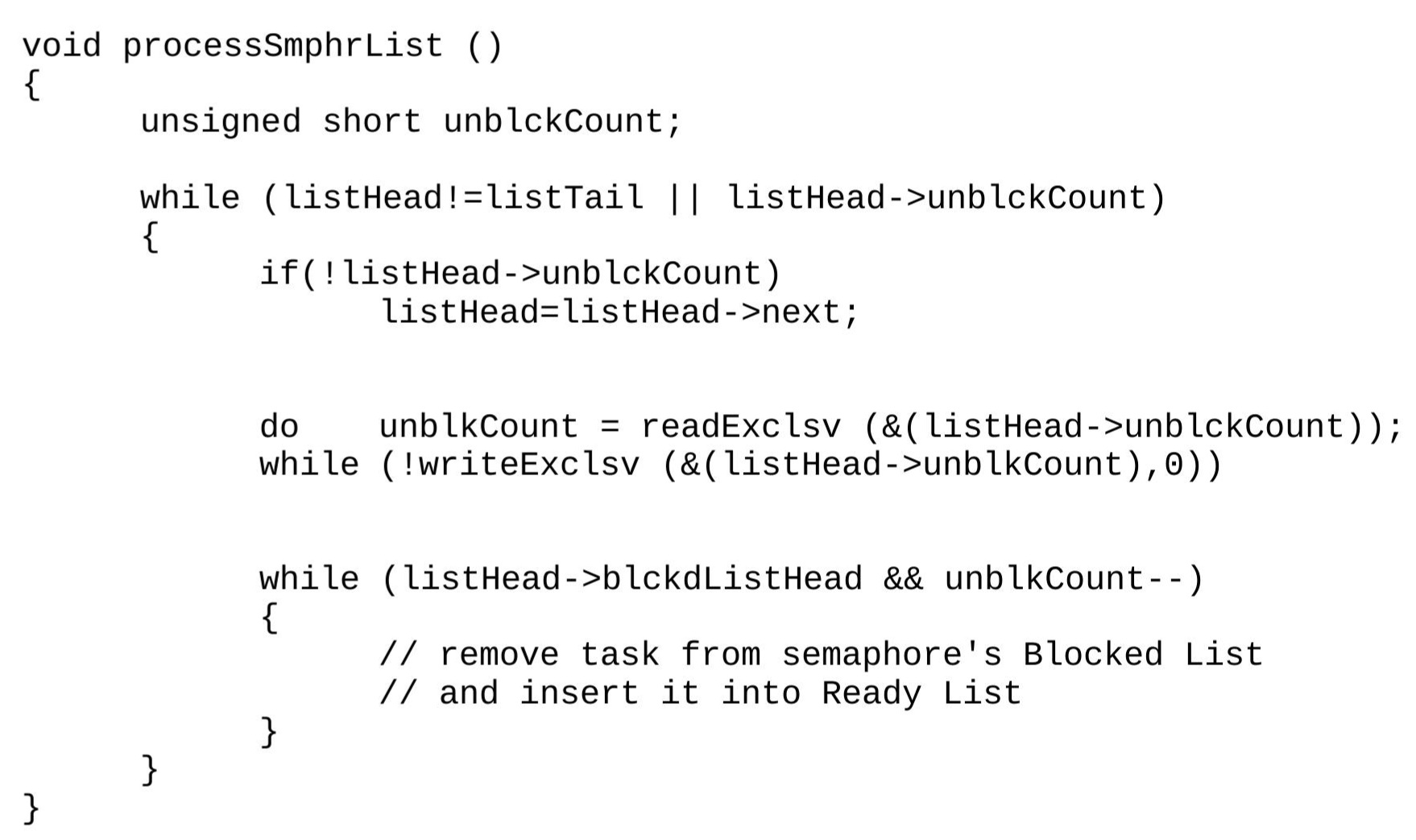}
  \caption{SW Interrupt Handler - Process FIFO Linked List}
\end{figure}
\subsection{Bitmap Flags}
This defer structure is unique from those previously presented in that it is not a FIFO structure.  The \enquote*{Bitmap Flags} variable is a word (32 bits in a 32 bit system) where each bit represents (is mapped to) a semaphore released from an interrupt handler.  If there are more than 32 semaphores released from interrupt handlers, then additional words can be used (herein assume NUM\_ISR\_SMPHRS\textless=32).
\hfill \break
\hfill \break
For an RTOS based on this defer structure, the number of semaphores released from interrupt handlers must again be statically specified in a configuration file.  Furthermore, however, the RTOS must declare an array of these semaphores during initialization.  When a semaphore is created at runtime, there must be a parameter to identify it as a semaphore released from an interrupt handler (dynamic configuration again necessary).  If it is, then in the RTOS function to create the semaphore, the next semaphore in the previously declared array is returned.  The index of such a semaphore in the array is also its bit position in the \enquote*{Bitmap Flags} variable (\&smphr - smphrArray).
\hfill \break
\hfill \break
Because each semaphore released from an interrupt handler has exactly one unique bit in the \enquote*{Bitmap Flags} variable associated with it, these semaphores must also contain an \enquote*{Unblock Count} variable (again only included for semaphores dynamically configured as semaphores released from interrupt handlers).  Therefore, as in an RTOS based on the FIFO Semaphores Buffer defer structure or the FIFO Linked List defer structure, two additional bytes are required for each semaphore released from an interrupt handler, and interrupt handlers must assert the software interrupt anytime that a semaphore is \enquote*{placed} in the defer structure.  See Fig. 6.  Atomically  updating the \enquote*{Bitmap Flags} variable (to \enquote*{place} a semaphore in the defer structure) is straightforward.  See Fig. 10.  The software interrupt handler should not loop through 32 bit positions.  In the worst case scenario, this would be O(n\textsuperscript{2}) and would impose unacceptable latency.  Instead, using binary search, a set bit can be discovered in 5 iterations.  In addition to atomically clearing the \enquote*{Unblock Count} variable, the software interrupt handler must also atomically clear the \enquote*{Bitmap Flags} variable.  An informal (uncompiled) simple example of how this is done is shown Fig. 11.
\hfill \break
\begin{figure}[h!]
  \centering
    \includegraphics [scale=0.237] {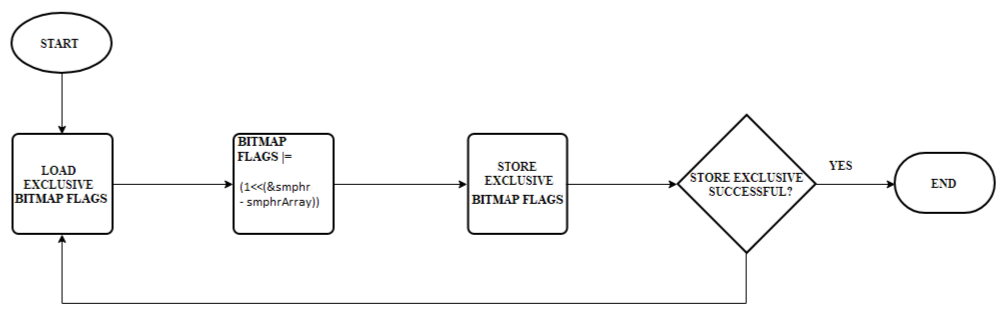}
  \caption{Atomic Instructions to Update Bitmap Flags Variable}
  \end{figure}
\begin{figure}[h!]
  \centering
    \includegraphics[scale=0.12]{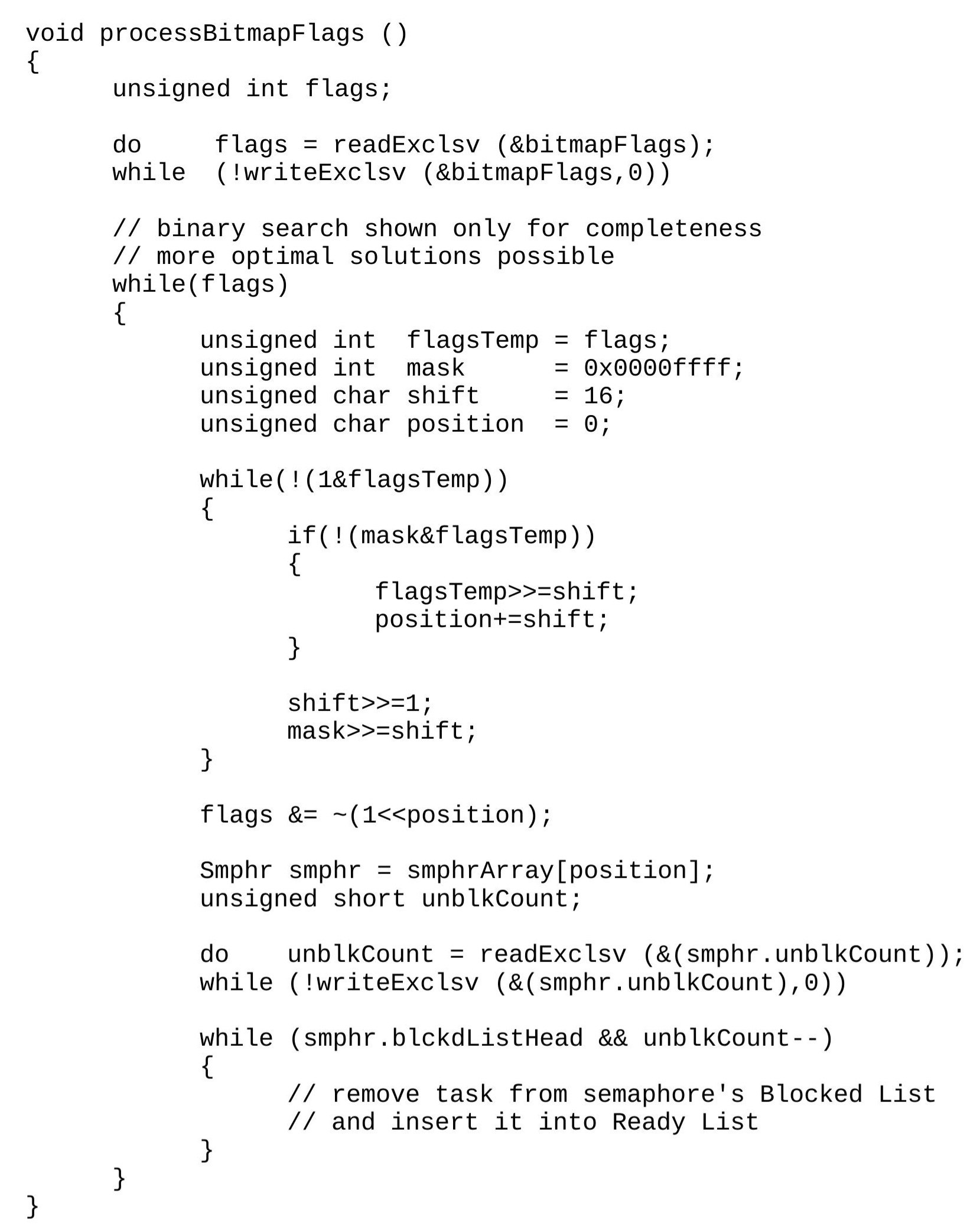}
  \caption{SW Interrupt Handler - Process Bitmap Flags}
\end{figure}
\hfill \break
\hfill \break
This type of defer structure has a minimal memory requirement as compared to the other defer structures: only 4 bytes (assuming NUM\_ISR\_SMPHRS\textless=32) for the \enquote*{Bitmap Flags} variable and 2 bytes for the \enquote*{Unblock Count} variable for each semaphore released from an interrupt handler.  The latency limitations, as compared to the FIFO Semaphores Buffer and the FIFO Linked List defer structures, are slightly worse as the unbuffering requires (a constant) 5 steps rather than one.
\hfill \break
\hfill \break
\hfill \break
\section{THE BARRIERS AND REQUESTS RTOS ARCHITECTURE}
The \textit{Barriers and Requests RTOS Architecture} is another approach to solving the \textit{Diminishing Bandwidth Problem}.  Interrupt handlers can directly unblock tasks.  This requires that they are not only able to atomically extract the highest priority task from a Blocked List, but also able to insert it atomically into the Ready List.
\hfill \break
\hfill \break
Note that interrupt handlers need only remove a task from a Blocked List and insert it into the Ready List.  They should not access the Delay List (of the \enquote*{Sys Tick} interrupt).  Rather, the state of a task that was blocked with a timeout (i.e. the task is in both a Blocked List and the Delay List with a state of \enquote*{Blocked Delayed} or similar) should be changed to \enquote*{Ready Delayed} before insertion into the Ready List.  The software interrupt handler or the \enquote*{Sys Tick} interrupt handler, upon encountering a task in the \enquote*{Ready Delayed} state will remove it from the Delay List and change its state to \enquote*{Ready}.  Not directly accessing the Delay List also serves to keep the length of the interrupt handlers short (although not as short as with the \textit{Deferred Structure RTOS  Architecture}).
\hfill \break
\hfill \break
Atomically extracting the highest priority task from a Blocked List is straightforward.  However, with regard to atomically inserting it into the Ready List, there are two approaches: A Sorted Ready List with an Atomic Insertion Algorithm or an Unsorted Ready List (which guarantees O(1) insertion and is straightforward to do atomically).
\hfill \break
In both cases, however, extracting the highest priority task from a Blocked List may be a problem if a task was in the middle of inserting itself into this Blocked List\footnote{Implied and included here is also a \enquote*{Sys Tick} interrupt handler that was in the middle of extracting a task from anywhere this Blocked List.}.  In this case, the inserting context will first set a \enquote*{Barrier} variable of the semaphore before the insertion begins.  This gives any potential interrupt handlers notice that accessing the Blocked List is not possible.  Interrupt handlers will rather then make a request (by atomically incrementing an \enquote*{Unblock Count} variable of the semaphore) to the inserting context, to unblock a task on their behalf.  The inserting context, after the insertion is complete, will then clear the \enquote*{Barrier} variable, and then will proceed to atomically unblock \enquote*{Unblock Count} number of tasks.  See Figs. 12, 13. Note that afterwards, if the inserting context determines that it has inserted itself at the head, and that the semaphore count is nonzero, it should reinsert itself into the Ready List.
\hfill \break
\begin{figure}[h!]
  \centering
    \includegraphics [scale=0.27] {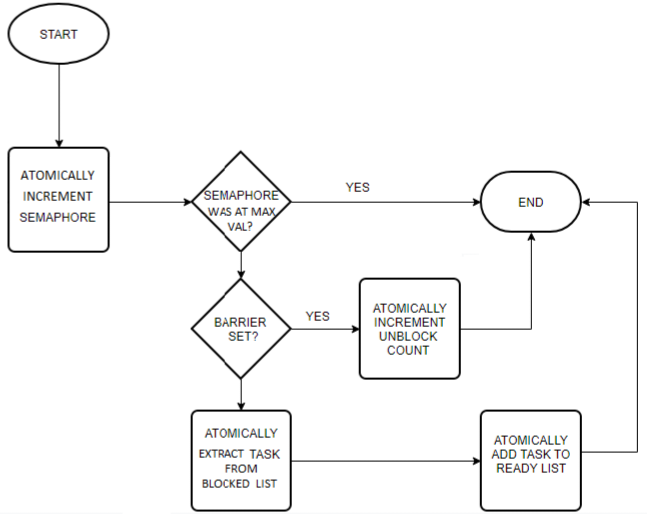}
  \caption{Barriers and Requests Interrupt Handler - smphrGvIsr() only}
  \end{figure}
\begin{figure}[h!]
  \centering
    \includegraphics[scale=0.27]{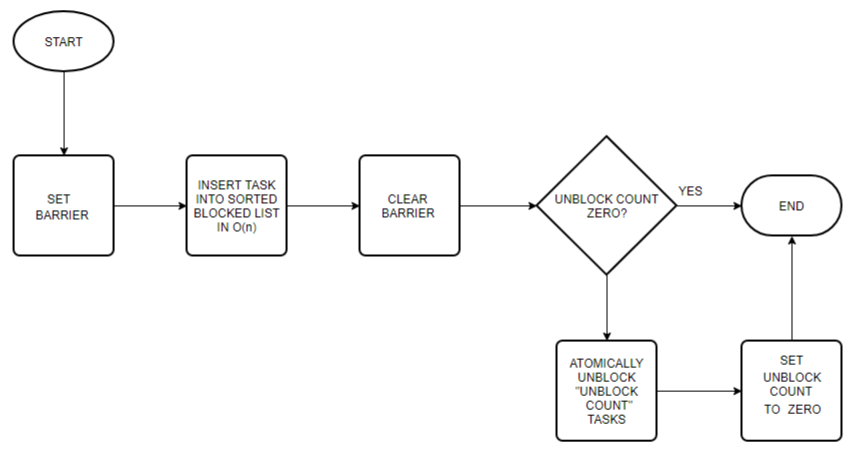}
  \caption{Barriers and Requests - Inserting into Blocked List}
\end{figure}
\hfill \break
\hfill \break
Similar to the FIFO Linked List defer structure, only dynamic configuration is required i.e. when a semaphore is created at runtime, an additional parameter is required to specify that this semaphore will be released from an interrupt handler; thus the semaphore will include the \enquote*{Barrier} and \enquote*{Unblock Count} variables.  The \enquote*{Barrier} and \enquote*{Unblock Count} variables, however, can be combined into one 2 byte variable without losing the maximum semaphore count value.  Simply, this \enquote*{Barrier And Unblock Count} variable is written to one by the inserting context, and if, upon exclusively reading it, an interrupt handler observes that it is not zero, it will increment it and exclusively write it.  After insertion, the inserting context will first read it exclusively and then write it to zero exclusively.  Then it will decrement the value that was read, and atomically unblock this number of tasks.
\hfill \break
\hfill \break
As in the \textit{Defer Structure RTOS Architecture}, when the currently executing task wishes to unblock a task or block itself, it must first disable the software interrupt (and the \enquote*{Sys Tick} interrupt).  In the case it is unblocking a task, it, like the interrupt handlers, must be able to atomically extract the task from the Blocked List and atomically insert it into the Ready List.  If the software interrupt is not disabled during this process, then a context switch may occur after the task has been extracted from the Blocked List but before it is inserted into the Ready List.  In the case a task is blocking itself, a context switch can render a Blocked List corrupt and inaccessible by interrupt handlers (due to the \enquote*{Barrier} variable remaining set).
\hfill \break
\hfill \break
\subsection{Sorted Ready List - Atomic Insertion Algorithm}
It is possible to insert atomically into a (singly linked) Sorted Ready List (assuming multiple inserters and only one lowest priority extractor).  The extractor, which only extracts at the \enquote*{head}, must do so atomically.  If an inserter (for example an interrupt handler) is preempted by multiple interrupt handlers, and multiple insertions take place, but all insertions occur in the part of the Ready List that the inserter has already traversed, then it is clear that this will not affect this inserter.  However, because the insertions could occur in the untraversed part of the Ready List, the inserter will always first read the \enquote*{next} pointer of the current list item exclusively.  Now, if the inserter determines that the insertion must happen at some point after the \enquote*{next} list item, then this fact will not change even if this inserter is preempted by multiple interrupt handlers and multiple insertions take place (before and after the \enquote*{next} list item).  Therefore, this inserter will eventually update its current pointer to that \enquote*{next} list item (even if it is no longer actually the \enquote*{next} item in the list) and repeat.  If, on the other hand, this inserter determines that it must insert right before the \enquote*{next} list item, and it is preempted by other interrupt handlers and multiple insertions take place, then if all the insertions take place after the \enquote*{next} list item, according to this inserter, there is no difference.  If, however, one or more insertions occur before the \enquote*{next} list item, then this inserter's attempt to exclusively update the \enquote*{next} pointer of the current list item will fail (something else was inserted) so it simply re-checks the \enquote*{next} list item (which will be different) by reading it exclusively.  Ultimately, this inserter, must execute a successful exclusive update to the \enquote*{next} pointer of the current list item to guarantee a successful insertion before the \enquote*{next} list item.  See Fig. 14.  If an interrupt occurring between a load exclusive and a store exclusive does not cause the store exclusive to fail (or if using a cmpxchg architecture), then it is clear that inserting at the \enquote*{head} must be done atomically.  Furthermore, however, when inserting right after the \enquote*{head}, one attempt must be made to atomically update the \enquote*{head} with its same old value.  This will ensure that the lowest priority extractor's attempt to atomically update the \enquote*{head} with the \enquote*{next} of the \enquote*{head} will fail.
\begin{figure}[h!]
  \centering
    \includegraphics [scale=0.27] {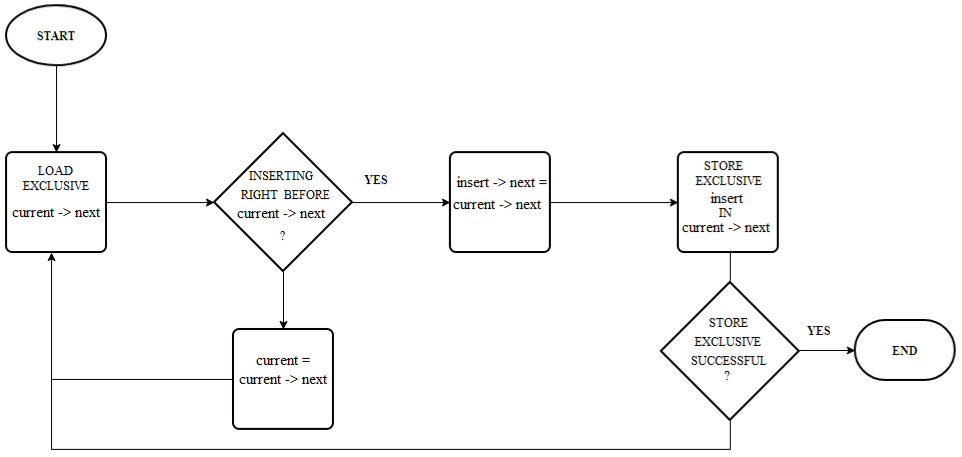}
  \caption{Sorted Ready List - Atomic Insertion Algorithm}
  \end{figure}
\hfill \break
Note that for a singular highest priority task, unblocked by a singular highest priority interrupt handler, this insertion is guaranteed to be O(1) (always at the \enquote*{head}).  Ultimately, all other tasks that receive their input from lower priority interrupt handlers should implement flow control at the application level (which, in the worst case, will effectively half the maximum bandwidth assuming each frame is acknowledged with an equal sized frame).  The reason flow control should be implemented, is because any such low priority interrupt handler can be preempted by at least one other entire interrupt handler, which will prevent it from being able to guarantee immediate retrieval of data from the associated peripheral
\hfill \break
\hfill \break
\subsection{Unsorted Ready List}
Interrupt handlers and tasks can insert into the Ready List atomically (at the \enquote*{tail}) if the Ready  List is unsorted.  For an unsorted linked list, the Insert operation is O(1) and the Extract Min operation is O(n) (exactly opposite to a sorted linked list).  When the highest priority task is unblocked, the software interrupt handler will find it at the \enquote*{tail} of the Ready List.  If all tasks are in the Ready List when this happens, the highest priority task's execution time will be increased by (8 cycles/iteration of the loop used to traverse the Ready List) * (n-1 iterations).  This additional latency may seem unacceptable but is actually negligible since the execution time of tasks is on the order of 1000s of cycles.  In fact, this latency is not definitively avoidable regardless of whether or not the Ready List is sorted.  For example, the software interrupt may be disabled (immediately before the highest priority task data becomes available) by a medium priority task that has unblocked the lowest priority task.  If all the tasks are in the Ready List, then the medium priority task must traverse the entire Ready List to insert the lowest priority task.  Thus, the highest priority task will suffer the same delay.
\hfill \break
\hfill \break
More importantly however, consider the Sorted Ready List case when immediately after the highest priority task is inserted into the Ready List, all other tasks are inserted into the Ready List in descending priority order.  This will increase the highest priority task's execution time by O(n\textsuperscript{2}).  Concretely, it would be (8 cycles/iteration of the loop used to traverse the Ready List)*(m\textsuperscript{2}/2+m/2) iterations, where m=n-1.  For an Unsorted Ready List, however, the increase would be only O(n).  In fact, this is another extensibility problem and a major deficiency of Sorted Ready Lists.  Therefore, the \textit{response time problem} shall henceforth be further specified to either be the \textit{bare metal response time problem} (associated with systems without an RTOS, as previously described herein), or the \textit{RTOS response time problem} (associated with any RTOS based on Sorted Ready Lists).
\hfill \break
\hfill \break
A simple optimization for an Unsorted Ready List (with one \enquote*{tail} used for insertion) is to have an Unsorted Ready List with k \enquote*{tails} (0 to k-1).  \enquote*{Tails} 0 to k-2 are dedicated for the insertion of tasks with priorities 0 to k-2 and Tail k-1 is for all of the rest of the lower priorities (an array of \enquote*{tail} pointers of size k should be declared and indexed using the priority of the task being inserted).  Such a Ready List will guarantee that the highest priority task (priority 0) can not only be inserted in O(1) but also extracted in O(1).  When traversing the lower priority tasks (inserted at \enquote*{tail} k-1), \enquote*{tails} 0 and 1 can be checked at each iteration to see if a high priority task has been unblocked.  Other similar optimizations are possible.
\hfill \break
\hfill \break
\hfill \break
\section{THE STRICTLY ATOMIC RTOS ARCHITECTURE}
It is possible for tasks accessing Blocked Lists to insert themselves (or the \enquote*{Sys Tick} interrupt handler accessing Blocked Lists to extract a task that has timed out from anywhere in the list) to do so without first setting a \enquote*{Barrier} variable.  Rather, another atomic insertion algorithm can be specified.  Again, this requires that the list be singly linked.  Furthermore, however, tasks must have one \enquote*{next} pointer for Blocked Lists and one \enquote*{next} pointer for the Ready List (and one \enquote*{next} pointer for the Delay List).  Note that since a task can never be in both a Blocked List and the Ready List, typically only two \enquote*{next} pointers are required.  Actually, typically, RTOS lists are doubly linked so tasks have four pointers (2 \enquote*{next} pointers and 2 previous pointers).
\hfill \break
\hfill \break
In the \textit{Strictly Atomic RTOS Architecture} interrupt handlers also, directly, and using atomic instructions, extract tasks from the \enquote*{head} of Blocked lists and insert them into the Ready List.  However, after extracting a task from a Blocked List, it's \enquote*{next} pointer (for Blocked Lists) is nulled out.  See Fig 15.  Tasks attempting to block (or the \enquote*{Sys Tick} interrupt handler attempting to extract from any position) can do so using only atomic instructions.  See Fig 16.  If an inserting task observes that the \enquote*{next} pointer of a task has been nulled out, it can conclude that this task is no longer in this Blocked List (it is, in fact, in the Ready List) and can restart at the \enquote*{head}.  Not shown is the case where the list is empty.  Also, since a null \enquote*{next} pointer can no longer denote a task at the end of the list, a task at the end of list is denoted by a \enquote*{next} pointer that points to the task itself.  And finally, note that afterwards, if the inserting context determines that it has inserted itself at the head, and that the semaphore count is nonzero, it should reinsert itself into the Ready List.
\begin{figure}[h!]
  \centering
    \includegraphics [scale=0.27] {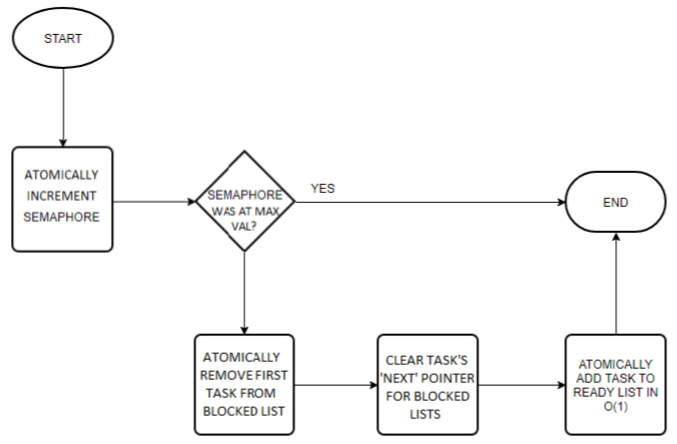}
  \caption{Strictly Atomic Interrupt Handler - smphrGvIsr() only}
  \end{figure}
\begin{figure}[h!]
  \centering
    \includegraphics [scale=0.25] {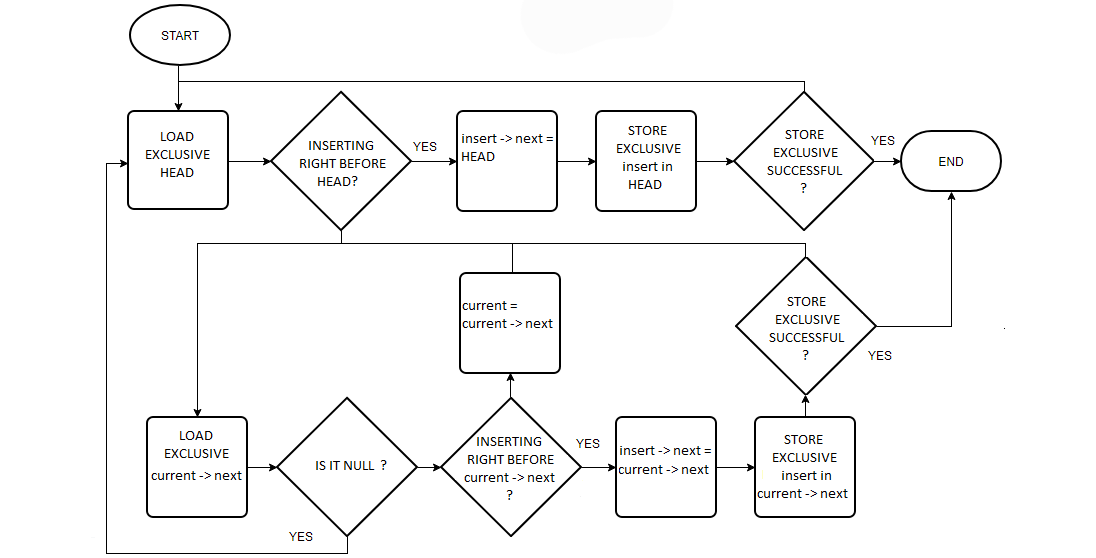}
  \caption{Sorted Blocked List - Atomic Insertion Algorithm}
  \end{figure}
\hfill \break
For this RTOS architecture, the values of the various \enquote*{next} pointers can be exploited to deduce what state a task is in (Blocked, Blocked Delayed, Ready, Ready Delayed, or Delayed etc).  Of course, this requires that a \enquote*{next} pointer is nulled out upon removing the task from the corresponding list.  Note that unlike all other RTOS architectures described, the \textit{Strictly Atomic RTOS Architecture} requires no additional memory.  In fact, since an RTOS typically uses doubly linked lists, and has two sets of \enquote*{next} and \enquote*{previous} pointers, and this architecture only has three \enquote*{next} pointers, it actually saves a word per task.  Singly linked lists, however, come with a disadvantage.  The Extract operation (extracting from a random position in the list) is O(n).  This introduces another version of the aforementioned \textit{RTOS response time problem}.  Consider that the highest priority task is running and all other tasks are blocked on the same semaphore.  However, they all have a timeout associated with the blocking i.e. they are all also in the Delayed List.  In the worst case, the \enquote*{Sys Tick} interrupt handler, will have to extract them all from the Blocked List (from the last task to the first task).  Only 16 tasks will impose 1000 more cycles latency upon the highest priority task.
\hfill \break
\hfill \break
\hfill \break
\section{HARDWARE SOLUTIONS TO THE DIMINISHING BANDWIDTH PROBLEM}
In the case that a functional RTOS is unavailable, there exist hardware techniques that solve the \textit{Diminishing Bandwidth Problem}.
\hfill \break
\hfill \break
\subsection{Without DMA Controller}
If a DMA controller is unavailable or unneeded, a GPIO pin can be sacrificed in order to guarantee that the incoming data of the highest priority task can never be lost.  The peripheral interrupt (of the peripheral that receives the highest priority task's data) is enabled, however its priority is set at a higher level than the highest priority acknowledged by the RTOS.  In other words, the peripheral interrupt handler cannot call any RTOS functions.  Therefore, it does not release semaphores or unblock tasks.  Furthermore, when interrupts are disabled by the RTOS, the peripheral interrupt is not disabled.  In the interrupt handler of the peripheral interrupt, a GPIO pin is flipped.  This GPIO pin has associated with it an interrupt that is of the highest priority acknowledged by the RTOS.  The GPIO interrupt handler, can and will release the semaphore of the highest priority task.  This technique solves the \textit{Diminishing Bandwidth Problem}, because even if interrupts have been disabled (by the RTOS), the peripheral interrupt will not be disabled, and can always execute and receive data from the peripheral.  In this case, this interrupt handler, instead of releasing a semaphore, will trigger the GPIO interrupt handler which will then release the semaphore and unblock the highest priority task.  If, at any time, more data is received at the peripheral, the peripheral interrupt handler is always guaranteed to not be disabled.  It is clear that the cost of this technique, is one GPIO pin.
\hfill \break
\hfill \break
\subsection{New DMA Controller Design}
The \textit{Diminishing Bandwidth Problem} can also be solved using a novel advanced DMA controller.  Current DMA controllers do not have the necessary functionality.  Current DMA controllers allow for the specification of the size of the buffer in memory to transfer data to, and the configuration of an interrupt to be asserted when a specified number of bytes have been transferred; or on the other hand putting the DMA controller into continuous circular mode with no interrupts.  Such DMA controllers can only solve the \textit{Diminishing Bandwidth Problem} if the incoming data is of fixed size.  However, if the incoming data comprises a header and payload, this creates a time constraint for the DMA interrupt handler to analyze the header and obtain the size of the payload.  If the number of cycles that interrupts can be disabled grows with the number of tasks, as previously discussed, this time constraint may be violated.
\hfill \break
\begin{figure}[h!]
  \centering
    \includegraphics [scale=0.237] {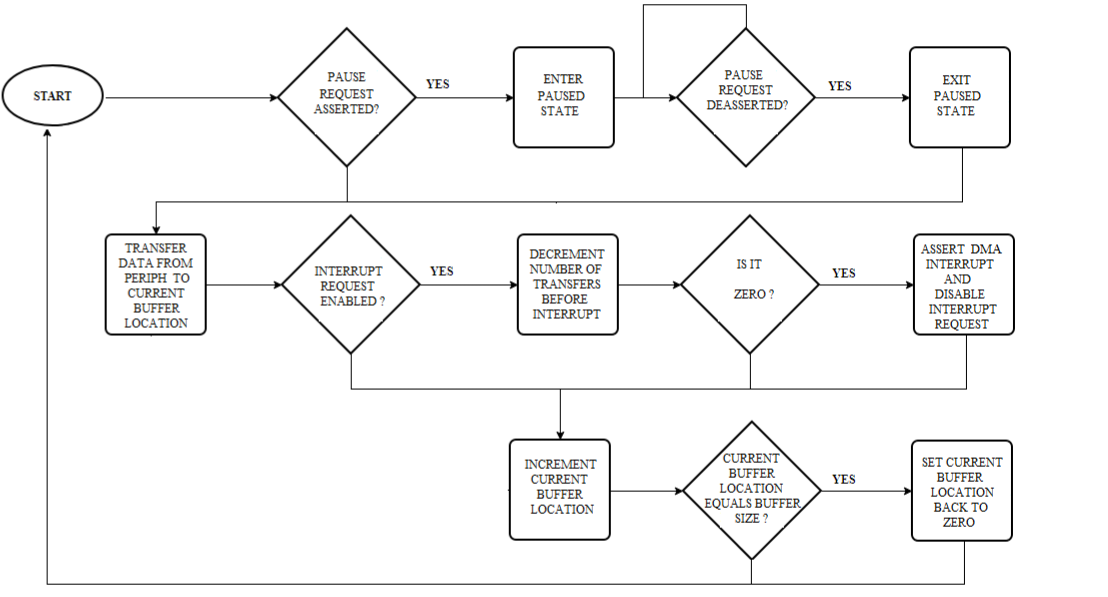}
  \caption{New DMA Controller Design}
  \end{figure}
\hfill \break
The novel DMA controller does not suffer from such limitations.  It can be configured in circular mode with interrupts, and allows for interrupts to be configured while the DMA is in the process of transferring data to the circular buffer.  In this way, even if the DMA interrupt handler is prevented from executing due to interrupts being disabled, an undetermined number of headers and payloads can still be continuously received in the circular buffer, and, ultimately analyzed by the DMA interrupt handler when it executes, such that even as the DMA controller is still transferring data, the DMA interrupt handler can update it with when the next interrupt should be asserted (based on the contents of the circular buffer).  Concretely, the DMA interrupt handler must be able to ensure that when it configures the DMA controller with when the next interrupt should be asserted, the DMA controller is not in the middle of changing state.  That is, when the DMA interrupt handler checks the current buffer location of the DMA controller (the current write location in the circular buffer) and updates it with the buffer location of the next interrupt, the DMA controller cannot be in the process of updating its current buffer location.  Hence there must be a novel write only \enquote*{pause request} field and a novel read only \enquote*{paused state} in the DMA controller.  The DMA interrupt handler will assert the \enquote*{pause request} field of the DMA controller and spin to check that it has entered the \enquote*{paused state}, and only proceed thereafter; and finally, when done configuring the DMA controller, it will de-assert the \enquote*{pause request} field.  The DMA controller will ignore pause requests if it is in the middle of transferring data and updating its status.  See Fig. 17.
\hfill \break
\hfill \break
\section{Multi-core Systems}
In single-core systems, the highest priority interrupt handler can never fail an atomic write operation due to a lower priority interrupt handler or a task that was attempting to atomically access the same memory location.  The reason is that the highest priority interrupt handler can never be preempted by lower priority interrupt handlers or tasks, thus guaranteeing that any atomic write operation occurs directly after the (previous) corresponding atomic read operation without impediment. However, for multi-core systems, the highest priority interrupt handler may actually fail an atomic write operation because lower priority interrupt handlers and tasks that are attempting to atomically access the same memory location, may be executing simultaneously on separate cores.  The constraint that the highest priority interrupt handler cannot be impeded must be maintained in multi-core systems.  Note that in general, multi-core systems exacerbate the  \textit{Diminishing Bandwidth Problem} because multiple cores can contend (via a spinlock) to execute O(n) operations with interrupts disabled).
\hfill \break
\hfill \break
In order to ensure that lower priority interrupt handlers and tasks running on other cores do not even cause the highest priority interrupt handler to fail an atomic write, the \textit{Defer Structure RTOS Architecture} must be used.  However, the semaphore value is not incremented in the interrupt handler.  See Figure 18.  This prevents contention by lower priority contexts executing simultaneously on other cores. Of course each core must have a separate defer structure and a separate 'Unblock Count' variable (that is never incremented past the maximum semaphore value).  The software interrupt handler will always process the defer structures of all cores (and atomically increment semaphore values).  Note that the FIFO Semaphore Counts Buffer can also be used for multi-core systems (no 'Unblock Count' variable).  In this case, however, the software interrupt must always be asserted after placing a semaphore reference in the Defer Structure.
\hfill \break
\begin{figure}[h!]
  \centering
    \includegraphics [scale=0.37] {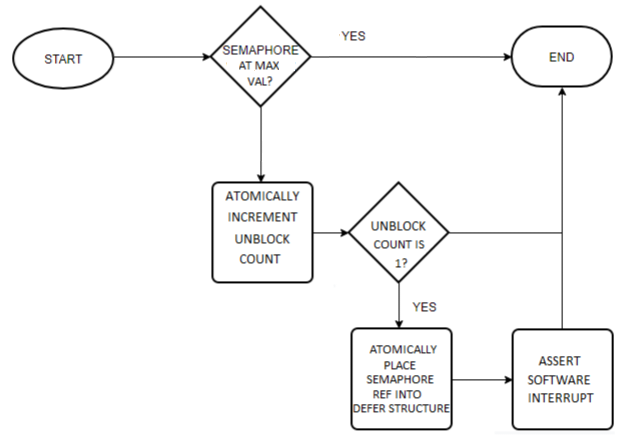}
  \caption{Multi-core Defer Structure Architecture}
  \end{figure}
\hfill \break
\hfill \break
\section{Conclusions}
Real-time operating systems were created to solve the bare metal response time problem: an extensibility problem whereby a lower priority task could prevent the highest priority task from, after receiving its input, producing its output within the deadline.  Unfortunately, this solution introduced a new extensibility problem - the \textit{Diminishing Bandwidth Problem} - whereby lower priority tasks and interrupt handlers, by traversing lists of tasks with interrupts disabled, could prevent the highest priority task from receiving its input in the first place.  In the worst case, this problem grows as more tasks are added to the system.  An RTOS should provide for total isolation of the highest priority task such that it cannot be impeded upon, in any way, by any other contexts.  Current RTOS architectures are dangerous in this regard and pose a serious risk to applications where human life or significant assets may be at stake.
\hfill \break
\hfill \break
\begin{table}[h]
\caption{RTOS ARCHITECTURES THAT SOLVE THE DIMINISHING BANDWIDTH PROBLEM}
  \centering
    \includegraphics [scale=0.15] {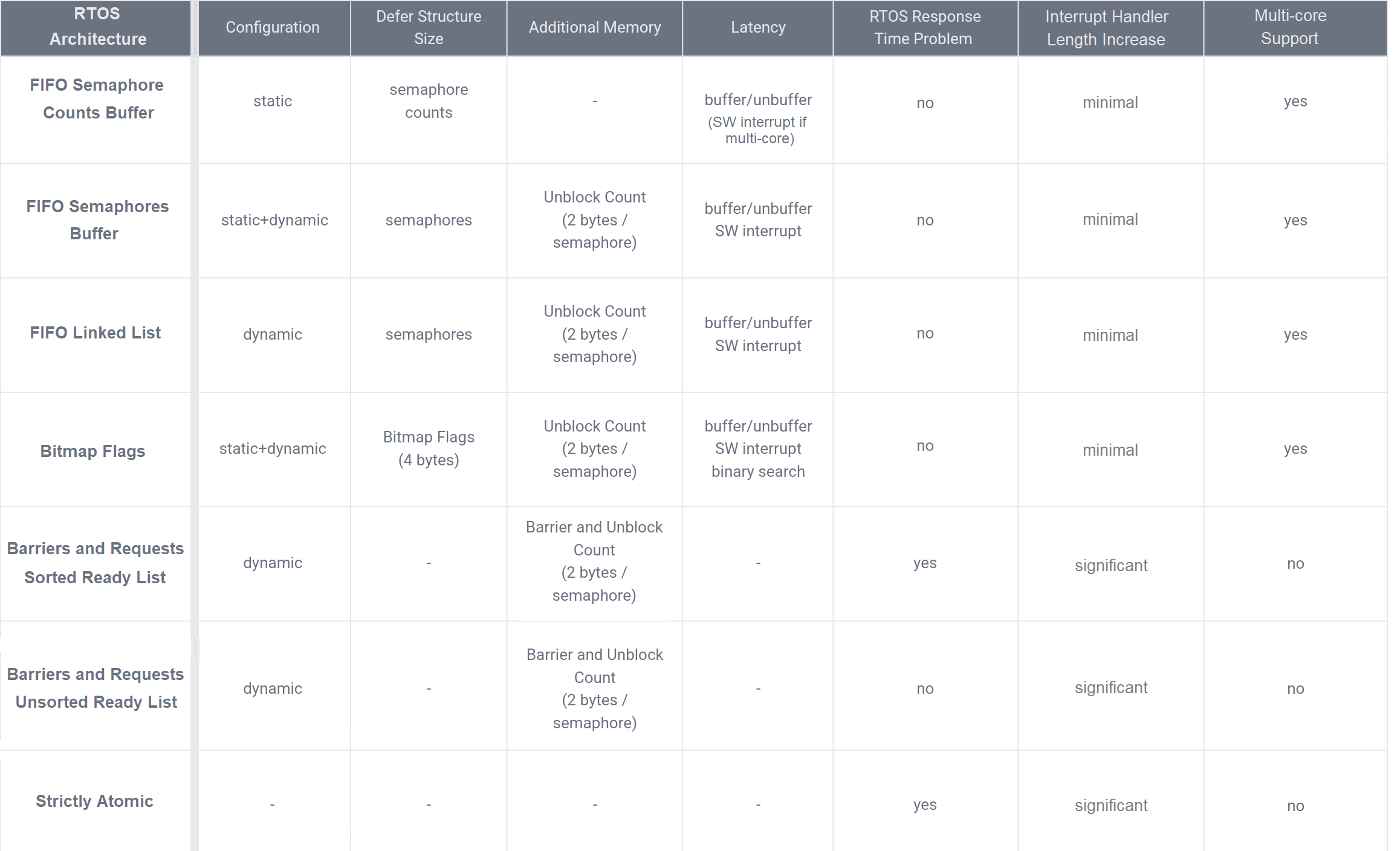}
  \end{table}
There are exactly three general RTOS architectures that will solve the \textit{Diminishing Bandwidth Problem}: the \textit{Defer Structure RTOS Architecture}, the \textit{Barriers and Requests RTOS Architecture}, and the \textit{Strictly Atomic RTOS Architecture}.  Different implementations of these architectures were described in detail.  Table I compares and contrasts these implementations.
\hfill \break
\hfill \break
The \textit{Defer Structure RTOS Architecture} is the optimal solution that is functional in multi-core systems, has no drastic increase in interrupt handler length, and avoids the \textit{RTOS response time problem} (so long as the Ready List is left unsorted).
\hfill \break
\hfill \break
\hfill \break
\hfill \break






\end{document}